# A robust and modular cesium magneto-optical trap using divergent laser beams

**Paola Luna[1], Kenneth Nakasone(中曽根ケネス)[1,*], Andrei Zhukov[1,*], Philip Wondrak[1], Brian P. Anderson[1], Benjamin F. Kanzer[1], Cristian D. Panda[1]**

[1] *Wyant College of Optical Sciences, The University of Arizona, Tucson, AZ 85721, USA*

[*] *The authors contributed equally to this work.*

Abstract: Magneto-optical traps (MOTs) are a workhorse technology for neutral-atom quantum sensing, simulation, and computing. Here, we demonstrate a MOT optimized for inertial quantum sensing and precision metrology applications. Our MOT traps $4\times10^8$ cesium (Cs) atoms in a robust, modular system that achieves stable operation through two design features: (1) we use divergent cooling laser beams to reduce unwanted reflections and (2) optical elements are mounted in a fiber-coupled, compact, and modular cage rigidly attached to the vacuum chamber. Following polarization-gradient cooling (PGC), the system produces atom samples with temperatures below 10 μK, similar to those achieved in conventional MOTs that use collimated laser beams. In addition, we observe trapping of $2\times10^7$ Cs atoms in a non-conventional MOT geometry, where the cooling laser beams are diagonal to the principal axis of the quadrupole magnetic field.

## I. INTRODUCTION AND MOTIVATION

Laser cooling and trapping have enabled a wide range of cold-atom metrology systems, including atomic clocks[1,2] and atom interferometers[3,4], with applications in precision measurements of fundamental physics[5-10] and inertial navigation[11]. In these systems, collecting atoms in a MOT is often the first stage of atom sample preparation[12]. The performance of the MOT directly affects the sensitivity and stability of the entire experimental system. Therefore, producing a cold-atom sample with consistently high atom number, low temperature, and robustness to environmental variations over months-long data taking campaigns is essential for both high-sensitivity laboratory measurements and in-field quantum sensors.

Conventional MOTs consist of three pairs of red-detuned, collimated, counterpropagating laser beams that overlap with the zero crossing of a quadrupole magnetic field[13]. This configuration is widely used but can also impose practical experimental constraints. Collimated beams passing through vacuum viewports can produce unwanted reflections that interfere and diffract, producing spatial gradients in the optical field[14] that affect MOT stability. In addition, the optical components required to control the alignment, intensity, and polarization of six independent beams can occupy substantial space around the vacuum chamber.

A wide variety of compact MOT geometries with reduced optical complexity using less than six trapping beams have been developed: pyramidal[15-17] conical[18], tetrahedral[19], on-chip[20-22] and grating[23-26] MOTs. These approaches show the importance of optical design in improving the practicality, stability, and scalability of cold-atom experiments. However, miniaturization of the optical system used to generate the cooling laser beams often comes with trade-offs in performance, such as reduced number of trapped atoms and higher temperatures. In contrast, large MOTs[27], where the atom number exceeds $10^8$-$10^9$, suffer from instabilities in number and spatial distributions of the atoms[28].

In this paper, we focus on building a compact tabletop MOT system designed to create a high-density MOT suitable for high precision quantum sensing applications, in our case interferometry with atoms trapped in an optical lattice[14,29]. Our MOT differs from conventional tabletop configurations in its use of divergent, rather than collimated laser beams[15,29-31]. A previous experiment[17] used highly-divergent laser beams to reduce the hardware footprint of a miniaturized MOT setup. In contrast, we employ divergent beams in our system to reduce instabilities in atom numbers and shot-to-shot variations in the spatial distribution of the atom sample typically caused by reflections from the vacuum chamber viewports. To our knowledge, our work is the first implementation of PGC with divergent laser beams, which results in temperatures of <10 μK, comparable to those obtained in collimated-beam MOT configurations.

We achieve a Cs MOT that traps $4\times10^8$ atoms with less than 20% fluctuation in atom number over a dataset taken for 50 hours. Additionally, each laser beam is built from optical elements assembled in a cage system that can be mounted directly on the vacuum chamber (Fig. 1(a)). This configuration preserves independent control of all laser beam properties (size, collimation, pointing, position, polarization, and intensity) while reducing space taken by beam delivery systems, improving mechanical stability, and providing quick assembly and disassembly for tabletop MOT systems.

Finally, we investigate a MOT configuration where the divergent beams propagate diagonally with respect to the principal axis of the magnetic-field gradient created by a separate set of quadrupole coils installed in our experiment (Fig. 1(c)). To our knowledge, a MOT with divergent beams in this diagonal configuration has not been previously reported. Using this configuration, we trap up to $2.6\times10^7$ Cs atoms and achieve temperatures of 10 μK. Although the MOT's long-term atom number stability is highly sensitive to small changes in experimental parameters like laser pointing and polarization, the observation of efficient trapping and cooling demonstrates the versatility of divergent-beam MOT geometries and motivates further exploration of alternative MOT configurations.

## II. EXPERIMENTAL SET-UP

A conventional MOT configuration uses three counterpropagating pairs of red-detuned, collimated laser trapping beams along each of the Cartesian axes to Doppler cool the atoms through velocity-dependent radiation pressure forces[13]. A pair of electromagnetic coils generates a quadrupole magnetic-field, whose gradient induces a position-dependent Zeeman shift of the atomic resonance along each axis. This Zeeman shift creates an imbalance in laser radiation pressure, resulting in a restoring force that localizes the atoms at the zero of the magnetic-field gradient.

We define a coordinate system referenced to the vacuum chamber (Kimball Physics MCF800-SphSq-G2E4C4A16) (Fig. 1(a)), where Z points upwards along the axis connecting the top and bottom ports. The Y coordinate connects the horizontal ports, opposite to the direction of travel of the atoms loading in the trap, and the X coordinate connects the centers of the 8" vacuum chamber ports, completing a right-handed coordinate system. A second coordinate system (denoted as prime) describes the axes of the images captured with a commercial camera (BASLER acA640-90um, Electrophysics 25 mm F/1.4 objective), with Z' approximately along Z, X' aligns with the direction the camera points towards the atoms. Lastly, the Y' coordinate completes a right-handed Cartesian reference frame.

Our MOT is loaded from a Cs atomic beam produced by a 2D MOT[27] inside a vapor cell connected to the experiment's main vacuum chamber, as shown in Fig. 1(a). Our 2D MOT laser delivery system consists of four counterpropagating divergent beams along the Z and X axes shaped by a combination of one plano-concave cylindrical mirror (Thorlabs LK1037L1-B) and two plano-convex cylindrical mirrors (Thorlabs LJ1309L1-B and LJ1567L2-B). The atomic beam formed by the 2D MOT travels to the main vacuum chamber through a differential pumping tube (diameter 5 mm, length 150 mm). This maintains a UHV pressure of $2\times10^{-10}$ torr (measured with Agilent UHV24p gauges) in the main chamber, much lower than the pressure inside the 2D MOT cell, 1.3×10$^{-6}$ torr Cs vapor pressure at room temperature. We observe a loading rate of the 3D MOT of 3×10$^{8}$ atoms/second, around 2-3 orders of magnitude higher than when the 2D MOT is disabled.

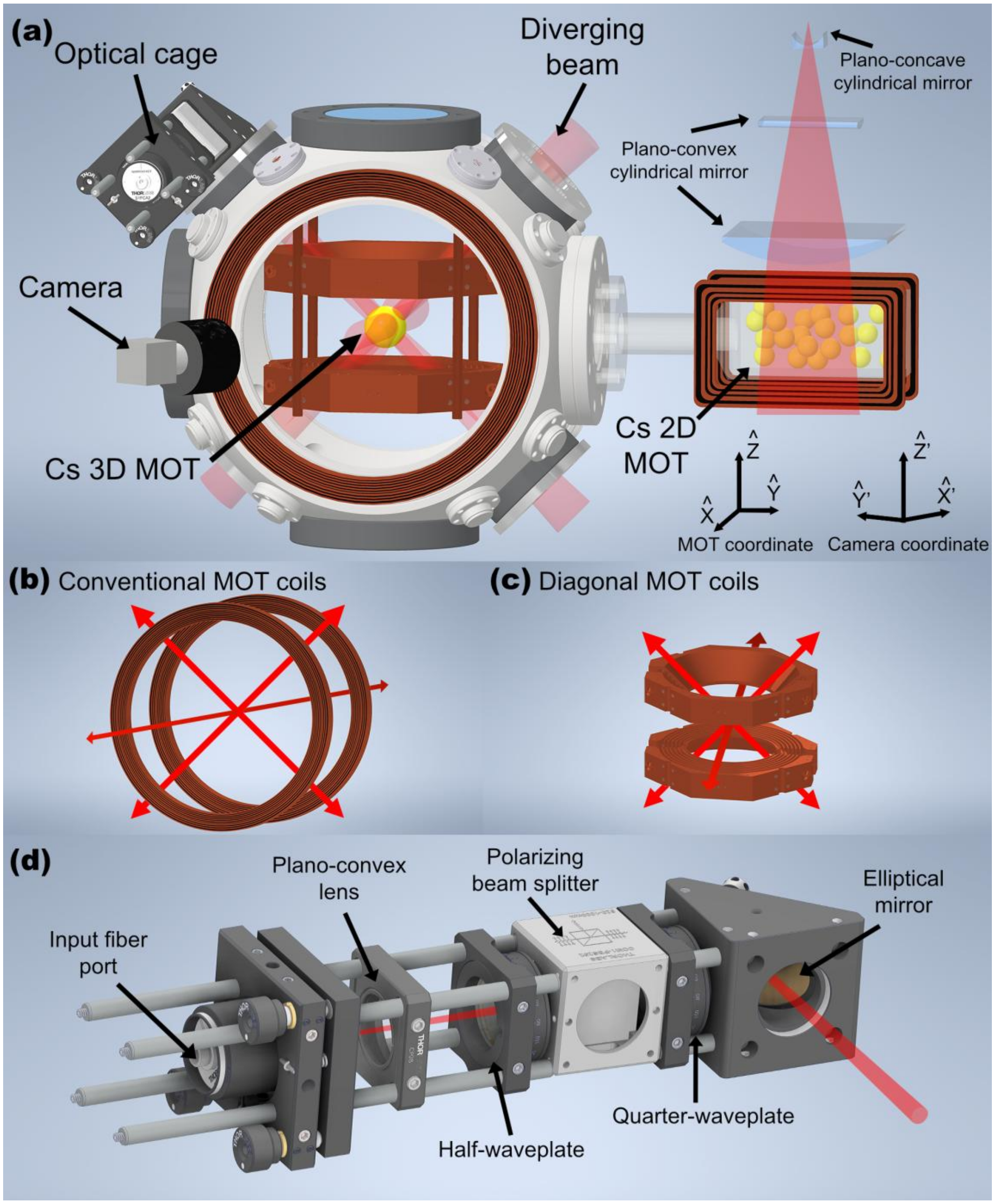

**Fig 1. Optical, magnetic and vacuum systems.** (a) The rectangular 2D MOT cell (right) contains Cs atoms at room temperature vapor pressure and is isolated from the primary vacuum chamber by a differential pumping tube. We create a 2D MOT with four laser beams (only the top one is shown) and a set of four electromagnet coils. The 3D MOT vacuum chamber (left) has four optical cages attached to each diagonal viewport (for simplicity, only the top-left cage is shown) and two optical cages on a horizontal breadboard for 3D MOT laser delivery (not shown). MOT laser beams (light red) enter the viewports and diverge by half-angle $\theta = 16$ mrad (beam divergence is exaggerated for illustrative purposes), which reduces the negative effect of reflections from the vacuum chamber windows on the stability of the MOT. A camera mounted outside the vacuum system is pointed at the Cs atoms at an angle between the + X and -Y coordinates. We create 3D MOTs using two different sets of electromagnetic coils: (b) shows the "Conventional MOT coils" where the principal axis of the quadrupole magnetic-field is aligned to one of the laser beams and (c) shows the "Diagonal MOT coils" where the principal axis of the quadrupole magnetic-field is aligned along Z, at 45° from two laser axes and 90° from the third. The latter coils were initially designed for levitating the atom sample during evaporative cooling. (d) The compact and modular delivery cage system provides position adjustment of all optical components.

In our MOT, the laser light contains two different tones: a "trap" component which is red-detuned by $\Delta_{Trap}$= $2\pi$×12 MHz = 2.3 Γ from the $6^2S_{1/2}$, F = 4 → $6^2P_{3/2}$, F′ = 5 cycling transition, where Γ = $2\pi$×5.2 MHz is the natural linewidth of the Cs D2 transition line, and a "repump" component, resonant with the $6^2S_{1/2}$, F = 3 → $6^2P_{3/2}$, F′ = 4 transition. Both lasers wavelengths are near 852 nm. The trap light is sourced from a distributed Bragg reflector laser diode (Photodigm 852.347DBRH-L-T08) that is beat-note-offset locked to a "reference" laser[32]. The reference is an External-Cavity Diode Laser (Toptica Eagleyard EYP-ECL-0852-00080-1500-BFW01-0005) stabilized to the transition $6^2S_{1/2}$, F = 3 → $6^2P_{3/2}$, F′ = 2 via modulation transfer spectroscopy with a Cs vapor cell[33]. The repump light is also derived from this reference laser. The repump and trap beams are combined and seed a tapered amplifier (TA) (EYE-TPA-0850-02000-4006-BTU02-0000).

After the TA, acousto-optic modulators are used to split the light and independently control the power coupled into fibers and delivered to the vacuum system for the 2D MOT and 3D MOT. The 3D MOT light feeds into a six-way fiber splitter (HaphiT FPCM-850-1X6-HP1-320-FC/APC) that distributes the MOT light into the six independent beam paths that enter the vacuum chamber. The 2D MOT light feeds a four-way fiber splitter (HaphiT FPCM-850-1x4-HP1-320-FC/APC) that distributes the 2D MOT light into the four independent beam paths that enter the Cesium vapor cell.

The magnetic field in our experiment is produced by a pair of anti-Helmholtz water-cooled electromagnetic coils, which we refer to here as "Conventional MOT coils" (Fig. 1(b)), made from enameled magnet wire mounted around the two largest vacuum chamber viewports centered around the X axis. This results in a quadrupole magnetic field with the principal axis aligned to one of the laser beams. Each coil is composed of 144 turns tightly wound in a circular mold potted in STYCAST 2850 epoxy. A copper tube is part of a water-cooling loop and wraps around the outer diameter of each coil to dissipate the generated heat. The typical magnetic field gradient that produces the position-dependent Zeeman shift required for MOT confinement is 6 G/cm. A separate pair of suspended in-vacuum coils, which we refer to as "Diagonal MOT coils" (Fig. 1(c)), are present in the apparatus for atomic levitation during future evaporative-cooling experiments[34]. We also apply a compensation magnetic field to null the laboratory's residual magnetic field with three pairs of external coils (referred to as "compensation coils") mounted on a frame around the vacuum system (not shown in Fig. 1) that produce uniform magnetic field components along the X, Y and Z axes, respectively.

The 3D MOT laser beam delivery system is designed to be compact, modular, robust against drift, and can be disassembled and reassembled with minimal need for optical realignment. We meet these design considerations with a system of six independent optical cages (Thorlabs 30 mm) assembled using commercially available parts as shown in Fig. 1(d). The MOT light is fed from the output fiber of the six-way splitter into an adapter plate (SM1FCA2). The beam diameter is controlled using an asphere (Thorlabs C340TMD-B) while the collimation is controlled with a plano-convex lens (Thorlabs LA1509) of 100 mm focal length. The polarization of the light exiting the fiber is matched to the transmission axis of a caged polarizing beam-splitter (PBS) (Thorlabs CCM1-PBS252) with a half-waveplate (HWP) (Thorlabs WPH10E-830) on a rotating mount. A quarter-waveplate (QWP) (Thorlabs WPQ10E-830) on a rotating mount converts the linear polarization into circular polarization, as needed for optimal MOT cooling. Finally, the beam reflects off an elliptical mirror (Thorlabs PFE10 M01) inside a right-angle kinematic mount (Thorlabs KCB1EC) and enters the vacuum chamber through a viewport. All components are connected by rods, which allow for quick adjustment of the distance between the cage optics.

Our laser beams have a Gaussian far-field half-angle divergence of $\theta = 16$ mrad. At the center of the chamber, where the MOT atom sample is loaded, the laser beams have an approximate 1/e$^2$ diameter of 15 mm. We obtained these parameters using the “razorblade method” to measure the width of the laser beams exiting the optical cages at a few positions along the propagation axis. For each position, we used a micrometer kinematic stage to carefully move a razorblade along the axis transverse to the laser beam. We used a lens to focus the optical power unblocked by the razorblade on a calibrated power meter. Each translation resulted in data of measured optical power vs. razorblade position, as shown in Fig 2(a). We extract the beam 1/e$^2$ diameter, $d$, for each razorblade measurement from the fit of the data to the complementary error function[35].

We then repeat the diameter measurement at varying positions along the propagation axis of the laser beam. We performed this measurement for two separate beam delivery cage systems: with the beam propagating in the -X̂ direction, and the beam propagating along the - Ŷ - Ẑ direction (Fig. 1(a)). We then fitted each dataset values to the far-field gaussian propagation equation $d(z) = 2 \cdot z \cdot \theta$ to extract the divergence angle, $\theta$, and beam diameter at the position of the MOT, $d(z = 137$ mm) (Fig. 2(b)).

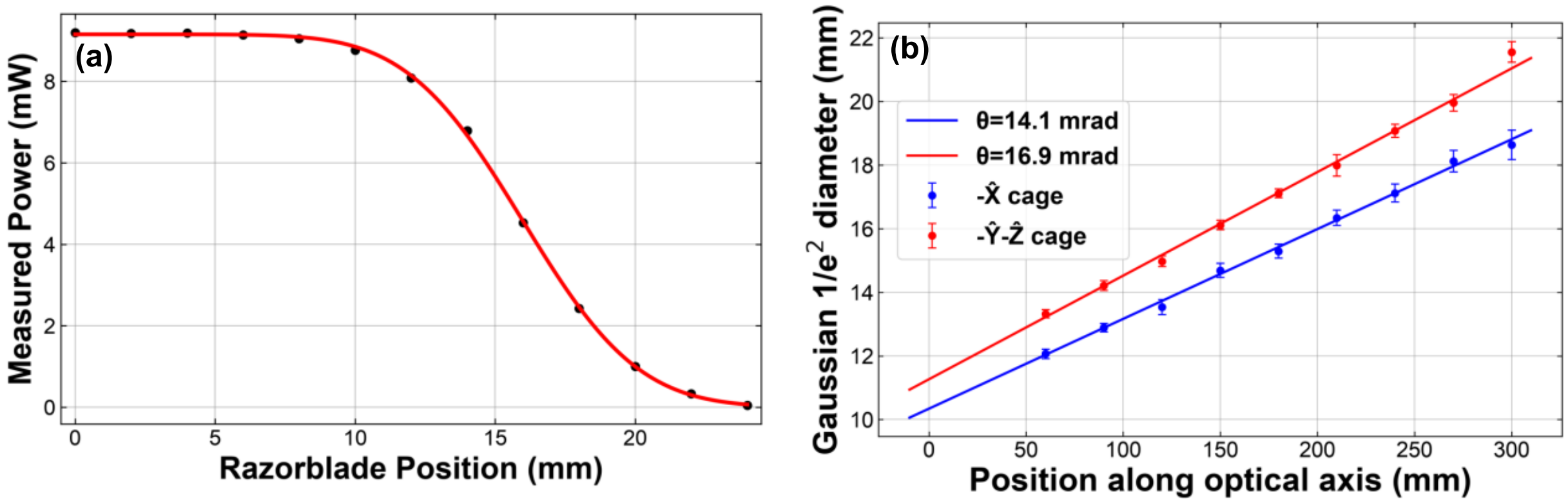

**Fig 2. Measurements of beam divergence angle.** (a) The power remaining after clipping the Gaussian laser beam with a razorblade is measured as a function of the razorblade position. A fit to the complementary error function (red) reveals the diameter of the laser beam, $d$. (b) We perform measurements of the beam diameter at various locations along the propagation axis ($z = 0$ is the position of the vacuum chamber window) for two different cages (with -X̂ and -Ŷ - Ẑ propagation directions). A linear fit gives the laser beam far-field divergence angles, $\theta$.

This geometry was chosen to reduce unwanted reflections from the vacuum chamber viewports, which have been previously observed to create parasitic interference that affects MOT stability[29,36]. The pointing and position of the laser beam are fully controlled by the kinematic mounts holding the fiber plate and the 45° mirror. The four cages that send light diagonally into the experiment are directly attached to the vacuum chamber, while the other two cages that send light along X sit on a horizontal breadboard at the center of the chamber. The MOT beams are aligned using custom-built transparent pinholes mounted on the vacuum viewports that define each optical axis. For each cage, the pointing is adjusted until the beam passes through both pinholes.

We set the laser-beam polarization to be circular, as required for best MOT performance, in the following way: each assembled cage was placed on the bench followed by a separate "calibration HWP" and "calibration PBS". The output power from this PBS is monitored with a photodetector. We then rotate both the QWP mounted inside the cage, or "cage QWP", and calibration HWP to minimize the transmission signal to the photodetector, which corresponds to the beam entering the calibration PBS having a clean linear polarization. We then add a "calibration QWP" between the cage QWP and the calibration HWP and rotate it until the PBS power outputs are balanced, indicating that the laser beam entering the calibration PBS is circularly polarized. We then rotate the cage QWP to either cancel out or double this polarization transformation and we record the two respective waveplate rotation angles: one where the PBS power output is minimized, the other when it is maximized, indicating cage output with circular polarization with the reference (or opposite) handedness, respectively.

## III. METHODS AND RESULTS

Each experimental shot follows a repeated timed sequence (using the open-source software Cicero Word Generator) where Cs atoms are loaded into the MOT, cooled by PGC, released for time-of-flight (TOF) expansion, and finally imaged. Each parameter was scanned with all the other parameters set to their optimized values.

### A. Imaging and temperature measurements

For imaging, all laser beams are switched off and the MOT sample is allowed to fall and expand for TOF (typically TOF = 5 ms), after which the MOT beams are pulsed on for $t_{image}$ = 0.1 ms (intensity $I_{Image}$ = 33.1 mW/cm$^2$ and detuning $\Delta_{Image}$ = $2\pi$×8 MHz). We choose $2\pi$×8 MHz to minimize scattering induced sample heating since we found no difference in detected fluorescence for detunings between 0 MHz to $2\pi$×10 MHz. We use a camera and objective to image the fluorescence photons emitted by the atom cloud. We then wait 50 ms for the atoms to fall out of the camera field of view, and we take a second image for background subtraction. A sample background-subtracted image of the MOT is shown in Fig. 3(a).

Each background-subtracted fluorescence image is analyzed in two ways. First, we integrate the camera signal over a rectangle that encloses the atom cloud and convert this signal into atom number using an independent calibration. For this calibration, a laser beam was split between a calibrated power meter and the camera. We used the measured power and the ratio between the two paths to calculate the conversion from camera signal to number of photons collected by the camera per unit time. We use a CAD model of the experiment to calculate the solid angle subtended by the camera collection optics system. After checking in a separate experiment that the atomic scattering is well saturated at the typical laser intensity, the atom-photon scattering rate is assumed to be $\Gamma/2$. The product of these three calibration factors gives the conversion from camera integrated fluorescence signal to atom number.

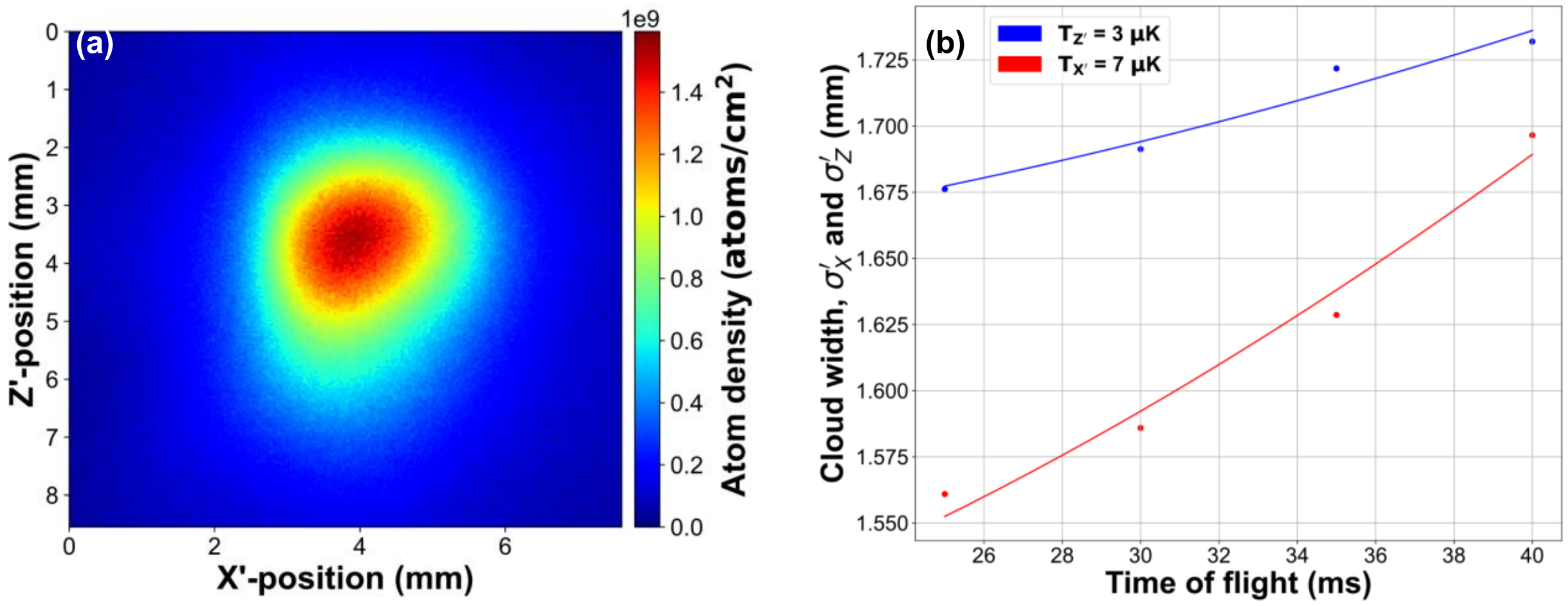

**Fig. 3. Temperature measurement procedure**. We capture a camera image of the MOT cloud (a), we fit Gaussians along the X' and Z' axes to find the corresponding widths of the atom sample. This is repeated at different TOFs and a dataset (b) is generated. We fit the data corresponding to each axis to Eq. (1) to compute the atom sample temperatures.

Second, we extract the Gaussian half-widths, of the atom-cloud spatial distributions, $\sigma_{X'}$ and $\sigma_{Z'}$ by fitting the data summed up along each axis to Gaussian functions. The temperature of the atoms, *T*, is measured using the time-of-flight method (TOF) by fitting the sample widths before, $\sigma(0)$, and after expansion, $\sigma$ (TOF), to the ballistic expansion equation:

$$\sigma^2(TOF) = \sigma^2(0) + (k_B T/m)(TOF)^2 \qquad (1)$$

where $k_B$ is the Boltzmann constant and *m* is the Cs atomic mass. A sample temperature fit is shown in Fig. 3(b).

Due to loading asymmetries, the shape of the MOT sample is not consistently Gaussian for TOF durations shorter than ~15 ms. Therefore, we use TOF durations longer than 20 ms to obtain consistent temperature measurements. The atoms fall outside of the camera imaging area after more than 45 ms TOF. We also note that variations in atom sample spatial distributions introduce significant uncertainty in the TOF method that limits accuracy when measuring temperatures below 10 μK, where other methods may be more suitable[37].

### B. Optimization of MOT parameters

Optimizing the efficiency and reproducibility of the MOT is an important step towards building reliable quantum sensing and metrology systems. We first optimize the loading of the MOT, where the main figure of merit is the number of atoms loaded. For MOT loading, we turn on the 2D MOT light, 3D MOT light, 2D MOT coils, and the “conventional 3D MOT coils” which together transfer Cs atoms from the 2D MOT into the 3D MOT. We perform a multi-parameter optimization of the 3D MOT loaded atom numbers by scanning the magnetic field gradient, the intensity, the duration of the MOT loading step, and the detuning of the MOT light (Fig. 4).

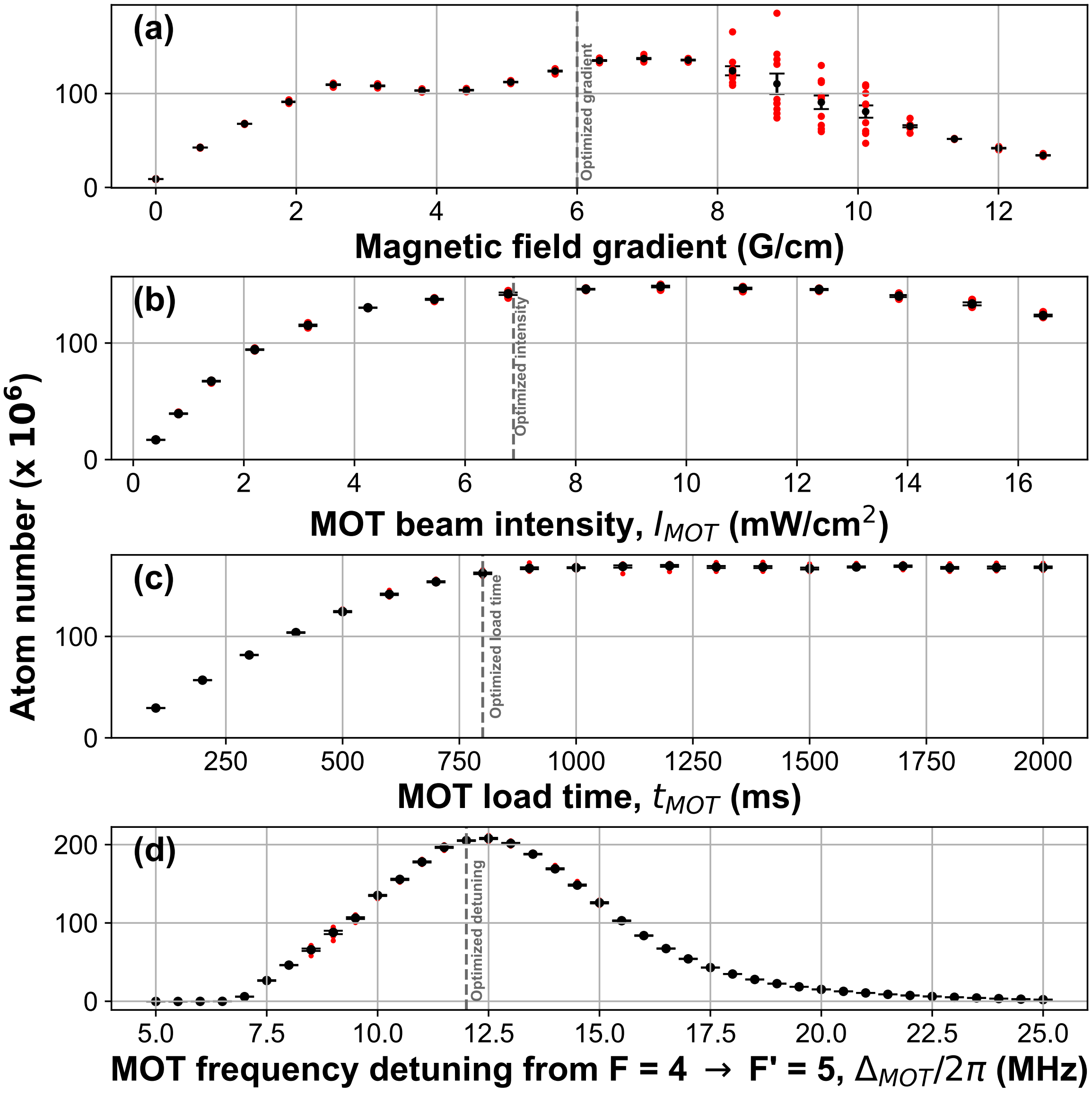

**Fig 4. Optimizing divergent-beam 3D MOT parameters for high atom number.** Scans of 3D MOT magnetic field gradient, intensity, load time and detuning. Each data point (black) was averaged over ten measurements (red dots). The error bars (black) indicate ±1σ, where σ is the standard error. The dashed vertical lines (gray) indicate the typical optimized values which were kept constant while scanning all other parameters.

Fig. 4(a) shows that the atom number increases with larger magnetic field gradients up to an optimum value of 6 G/cm. We observe that higher values of the magnetic field gradient cause the MOT to become unstable, resulting in large shot-to-shot fluctuations and decreased atom number, possibly due to higher sensitivity to laser beam and magnetic field alignment. Fig. 4(b) shows that the atom number reaches a maximum at intensities near $I_{Trap}$ = 6.8 mW/cm$^2$, a well-studied behavior[38]. Fig. 4(c) shows an increase in the atom number trapped with $t_{MOT}$, which plateaus after a load time of approximately 800 ms. Fig. 4(d) shows that fewer atoms are captured at small laser detuning because of higher photon scattering and heating. At larger detunings, we see a drop in the atom number from inefficient loading due to low photon scattering. We find the optimized frequency detuning to be $\Delta_{Trap}$= $2\pi\times12$ MHz or 2.3 Γ. In addition to the parameters shown here, we have also optimized the intensity and detuning of the repump laser ($I_{Repump}$ = 0.75 mW/cm$^2$ and $\Delta_{Repump}$ = $2\pi\times0.5$ MHz).

The total number of atoms trapped in the MOT is limited by collisional and radiation pressure effects. We observe close to 1 billion atoms when using low laser beam intensity and low magnetic field gradients that produce a relatively large, cm-scale trap volume. However, in the data shown above we have also optimized for high atom number density, which typically results in trapping several hundred million atoms in mm-sized trap volumes. These atom numbers are comparable to other Cs MOTs that use conventional collimated, rather than divergent, six-beam configurations[39,40].

**C. Optimization of PGC parameters**

Sub-Doppler cooling in Cs MOTs is known to be sensitive to optical and magnetic-field parameters. Previous measurements with six-beam geometries[41-43] show that PGC can produce μK-scale temperatures when experimental magnetic fields and laser beam parameters are properly controlled. In our system, we implement a PGC cooling stage after loading the MOT. The quadrupole field is switched off, and the compensation coils are used to null the laboratory's magnetic field. To minimize the temperature of the atom sample during the PGC process, we scan the following parameters: beam intensity, frequency detuning, and PGC duration (Fig. 5).

Fig. 5(a) shows a sharp decrease in temperature when increasing laser beam intensity up to an optimum of $I_{PGC}$ = 1.5 mW/cm$^2$. In Fig. 5(b), we scan the laser detuning for three different values of laser beam intensity and observe the optimum trap laser detuning to be $\Delta_{PGC}$ = $2\pi\times120$ MHz for the lower intensity of 0.7 mW/cm$^2$. We observe that larger detunings minimize the atom temperature for higher laser beam intensities, consistent with higher photon scattering rates. Scanning the duration of the PGC process in Fig. 5(c) shows an optimum PGC duration of $t_{PGC}$ = 6 ms, which minimizes the temperature in both the camera X' and Z' axes by balancing short-term cooling with long-term heating by incoherent photon scattering.

The PGC process requires nulled magnetic fields to minimize the atom temperature. Fig. 6 shows scans of atom temperature versus magnetic field applied with the compensation coils. We observe a strong dependence of atom sample temperature on the applied field, with best cooling performance at values where the compensation coils null the laboratory's residual fields. This follows previously demonstrated behavior for PGC[43].

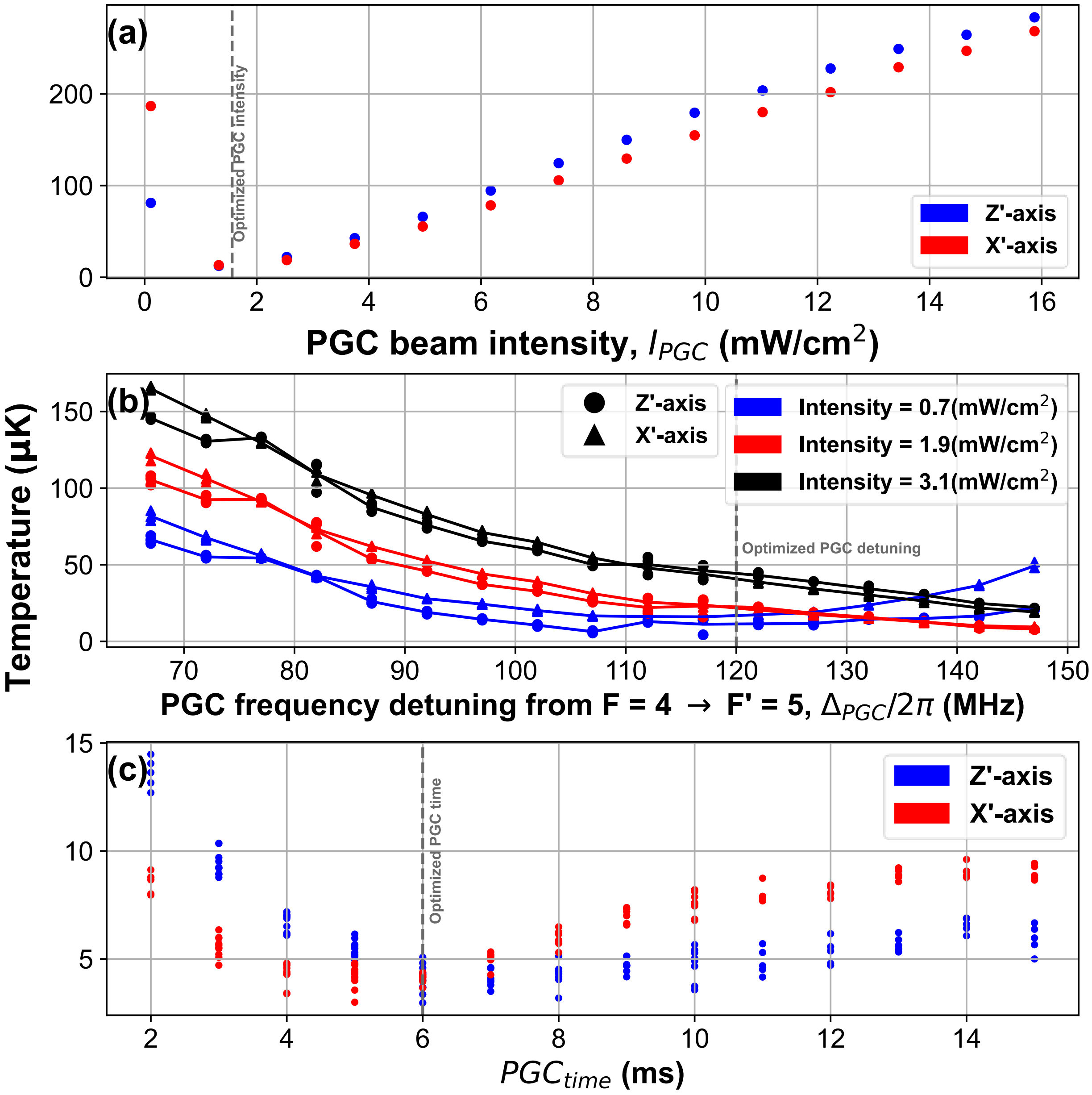

**Fig. 5. Optimizing divergent-beam MOT parameters during PGC to achieve low temperatures.** Scans of PGC beam intensity, frequency detuning, and time. The dashed lines (gray) indicate the optimized values chosen for each parameter, which were kept constant while scanning all the other parameters.

The temperatures obtained are comparable to those achieved by conventional MOTs that use collimated laser beams, which confirms the PGC process is effective in cooling MOTs that use divergent laser beams.

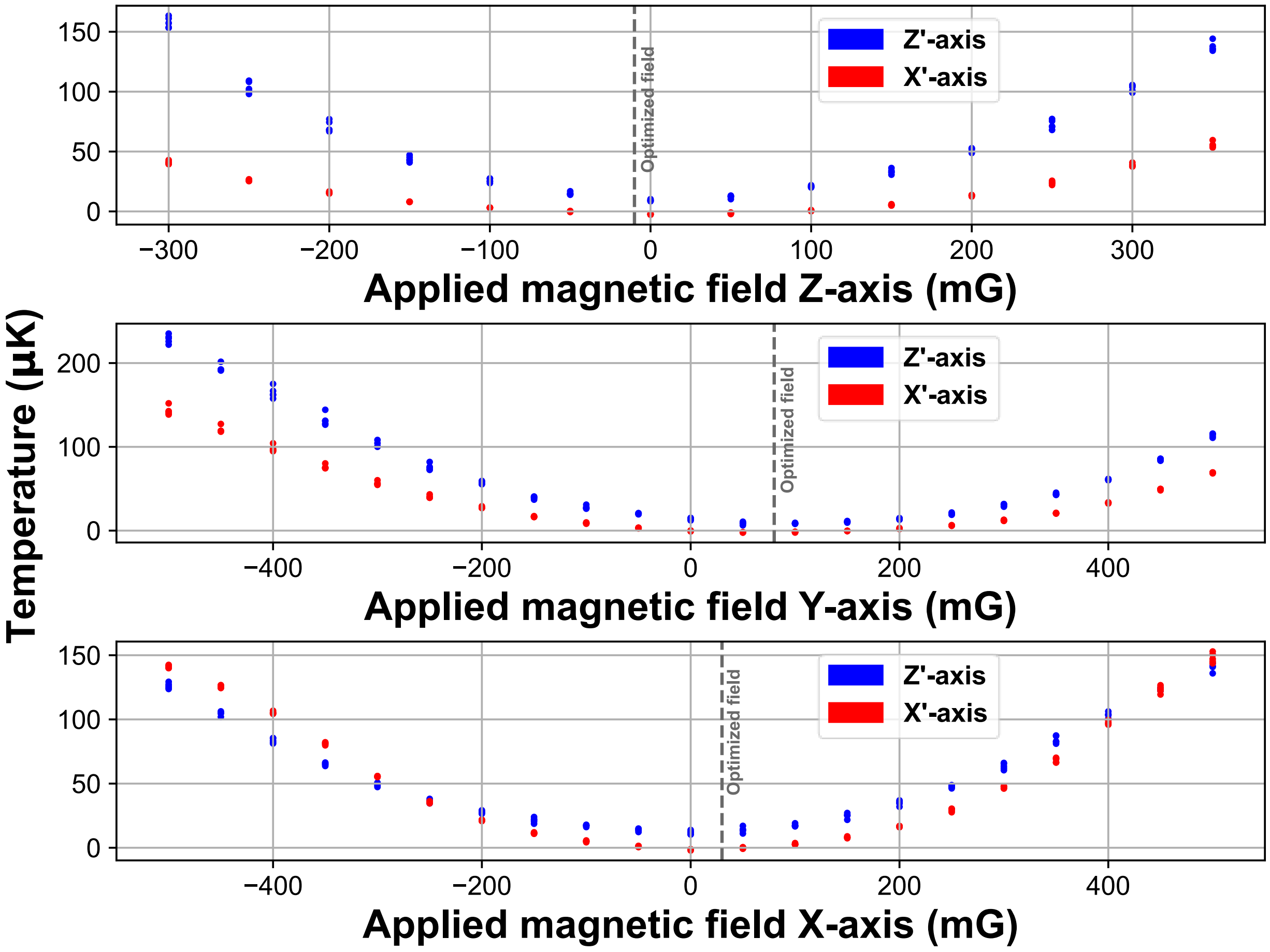

**Fig. 6. Optimization of atom sample temperature after PGC vs. applied compensation magnetic field.** Magnetic fields along the X, Y, and Z axes are varied by changing the current of each corresponding compensation coil to optimize atom sample temperature after PGC. Each parameter was scanned with all the other parameters set to their optimized values.

### D. Long-term MOT atom number and temperature stability

Consistent, long-term operation of the experimental system is critical for applications in fundamental gravitational physics, where small signals must be averaged over long integration times. Furthermore, quantum sensing systems must remain stable despite large variations in environmental parameters such as temperature and humidity. We characterize the stability of our divergent-beam MOT by measuring the temperature and number of trapped atoms over more than 50 hours (Fig. 7).

The MOT remains operational throughout this period, with temperatures remaining below 10 μK along both X' and Z' axes. The atom number is stable to within 20%. The atom-number fluctuations are correlated with variations in the lab temperature of up to 5 ℉ caused by a cold weather front passing through Tucson, Arizona during the mid-May data-taking period. In previous tests, such changes in temperature and atom number have been correlated with drift in the polarization output of the six-way distribution fiber splitter. This effect can be mitigated by temperature stabilizing the fiber splitter. Overall, our data show that we can operate the experiment over prolonged hours with minimal long-term fluctuations in atom number and temperature.

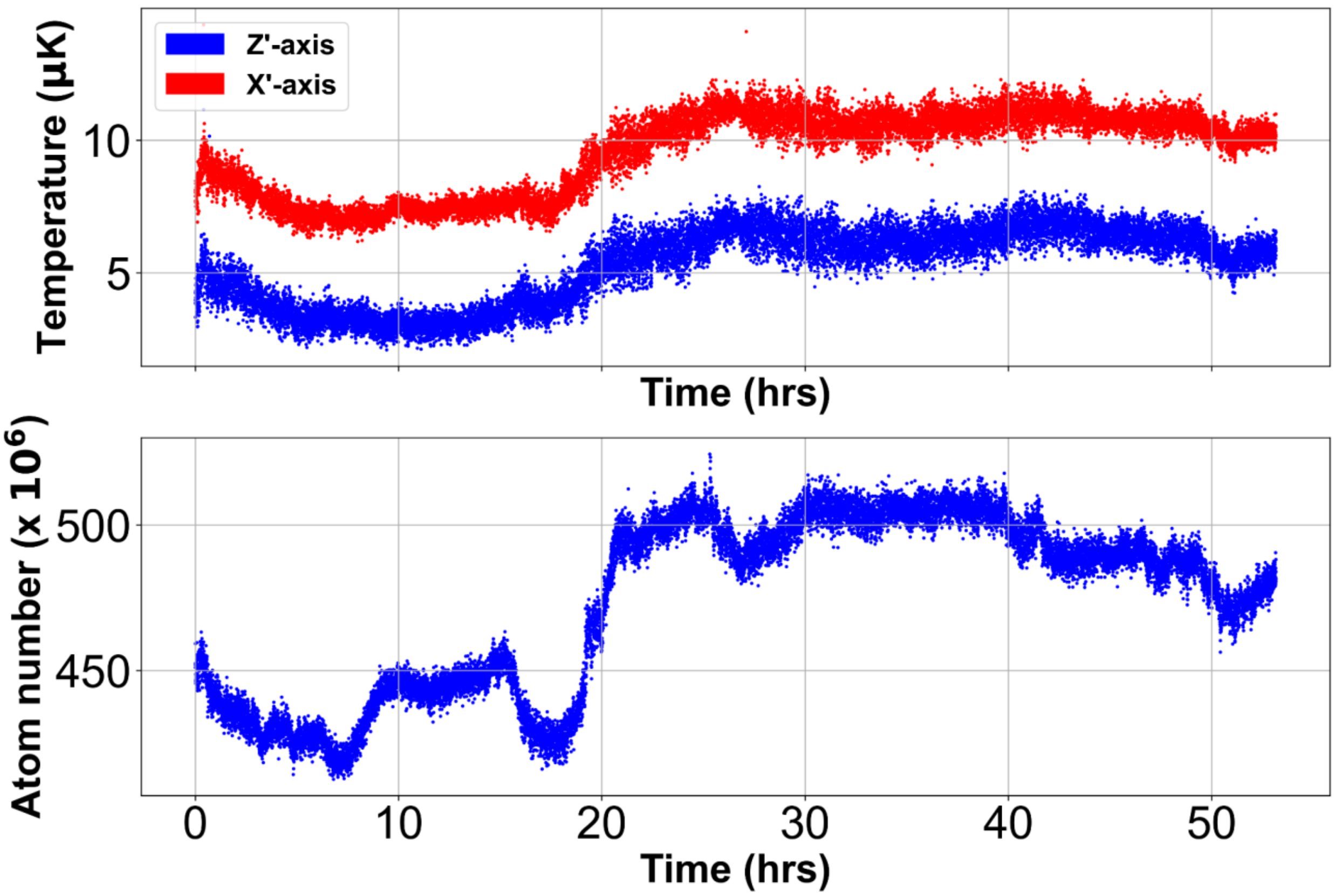

**Fig. 7. Long-term measurement of divergent-beam MOT atom sample stability.** Variations in the atom-sample temperature and number are measured over a period of over 52 hours.

**E. Diagonal-beam MOT configuration**

Unlike the more conventional MOT geometry discussed above, the first MOTs in our experiment were created in a configuration in which the principal axis of a quadrupole magnetic field is diagonal to the propagation axes of the laser beams. In this configuration, the confinement field was generated by the pair of in-vacuum "Diagonal MOT coils" shown in Fig. 1(c), which were initially developed for magnetic levitation of the atoms during a planned future implementation of evaporative cooling[30]. These coils were first installed in the vacuum system before installing the set of coils that we used for creating the MOT discussed in the previous sections of this paper ("Conventional MOT coils" in Fig. 1(b)).

In this diagonal-beam MOT geometry, the propagation axes of four trapping laser beams are at approximately 45° relative to the principal axis of the quadrupole field. As a result, the trapping force depends spatially on the relative angle between the magnetic-field gradient and the laser beam. We have created diagonal-beam MOTs with different spatial distributions by tuning various parameters, such as magnetic-field gradient, laser beam intensity and detuning. These MOTs exhibit high variance in their shape as these parameters are tuned, as observed in Fig. 8.

Fig. 8(a) shows the optimized MOT sample. In general, we observe high variability in atom number and shape over time, although the number stays stable around $2.6\times10^{7}$ atoms over a few minutes. We measured temperatures of approximately 13 μK along both X' and Z'. This demonstrates that PGC is still effective at reducing the atom sample temperature in this configuration.

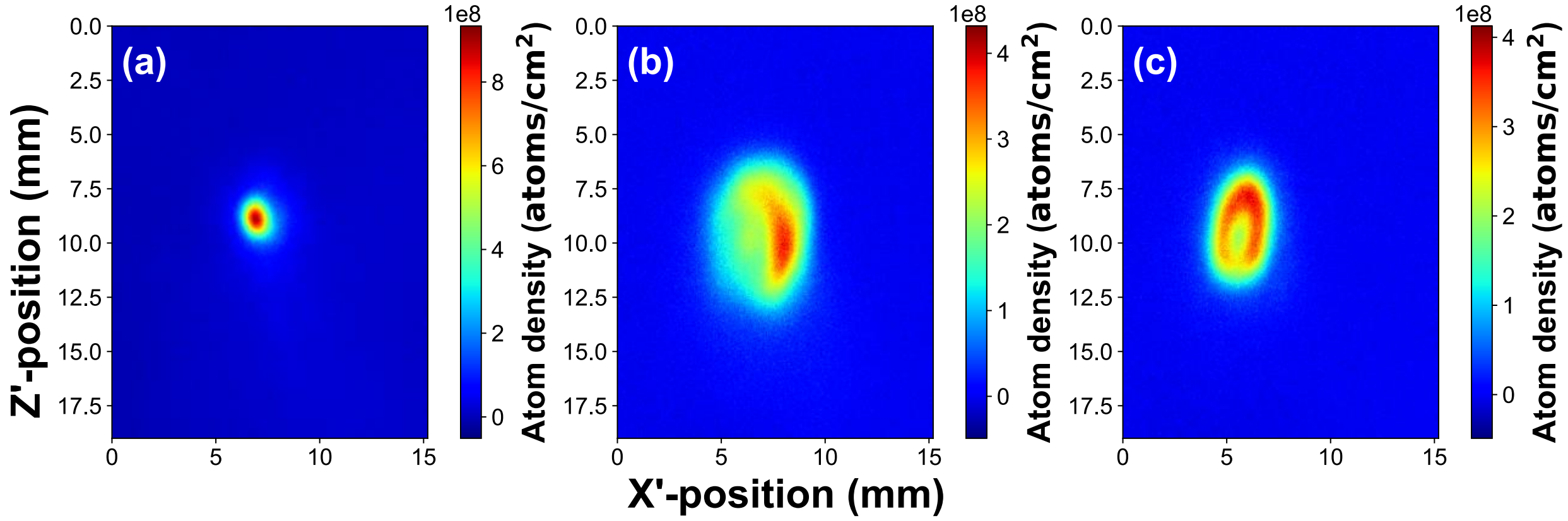

**Fig. 8. MOT produced in diagonal-beam configuration.** (a) Shows a high-density atomic MOT after optimizing the laser beam intensity, detuning and magnetic-field gradient, while (b) and (c) show various MOT shapes observed while changing these parameters.

These experiments demonstrate that it is possible to create a MOT where laser beams are diagonal to the principal axis of the quadrupole magnetic-field gradients. However, we observe that this configuration is much more sensitive than the conventional MOT configuration to changes in the polarization and pointing of the laser beams. We observed that drift in these experimental parameters causes a reduction in atom number and changes in the spatial distribution of the atom sample over timescales of tens of minutes.

## IV. CONCLUSION

Here, we have described an implementation of a Cs MOT that achieves long-term stability in atom number and temperature at the 20% level by using (1) divergent laser beams to suppress reflections from the vacuum chamber and (2) a compact, robust, and modular cage system to deliver laser beams to the atom sample. We have optimized the MOT atom number and temperature by tuning beam intensity, frequency detuning, magnetic field gradient, and loading time. We achieve samples with several $\times 10^8$ atoms at sub-Doppler temperatures of less than 10 μK. We also demonstrated a diagonal-beam MOT configuration, which can create atom samples with more than $10^7$ atoms and temperatures near 10 μK. However, this configuration is more susceptible to long-term drift in experimental parameters. The resulting Cs MOT is suitable for atom-interferometric applications in precision quantum metrology and quantum inertial sensing.

## ACKNOWLEDGMENTS

We thank Poul Jessen, Jason Jones, and Jae Hoon Lee for helpful discussions; Charles Condos, Dalziel Wilson, and Stanley Poul for sharing equipment; Richard Schickling, Paul Schickling, Jerry Edwards (Universal Cryogenics), and Brayden James for technical assistance; Stone Guymon and Scarlett Bolton for contributions to experiment assembly; Justin Murante for software support.

This material is based upon work supported by the University of Arizona. P. Luna is supported by the National Science Foundation Graduate Research Fellowship Program under Grant No. DGE-2137419. Any opinions, findings, and conclusions or recommendations expressed in this material are those of the author(s) and do not necessarily reflect the views of the National Science Foundation.

## AUTHOR DECLARATIONS

### Conflict of Interest

The authors have no conflicts to disclose.

### Author Contributions

**Paola Luna:** Data Curation (lead), Formal Analysis (lead), Investigation (lead), Methodology (supporting), Project Administration (supporting), Visualization (lead), Writing/Original Draft Preparation (lead), Writing – review and editing (equal). **Kenneth Nakasone:** Investigation (lead), Methodology (supporting), Project Administration (supporting), Resources (equal), Validation (equal), Writing – review and editing (equal). **Andrei Zhukov:** Investigation (lead), Resources (equal), Validation (equal), Writing/Original Draft Preparation (supporting), Writing – review and editing (equal). **Philip Wondrak**: Data Curation (lead), Investigation (supporting), Writing – review and editing (equal). **Brian P. Anderson**: Conceptualization (lead), Writing – review and editing (equal). **Benjamin F. Kanzer**: Investigation (supporting), Writing – review and editing (equal). **Cristian D. Panda**: Conceptualization (lead), Investigation (lead), Methodology (lead), Project Administration (lead), Supervision (lead), Writing/Original Draft Preparation (supporting), Writing – review and editing (equal).

### Data Availability

The data that support the findings of this study are available from the corresponding author upon reasonable request.